\documentclass[10pt,twocolumn,astron]{article}  	
\usepackage[margin=1in]{geometry}                		
\usepackage{amssymb}

\usepackage{amsmath}
\usepackage{amssymb}
\usepackage{empheq}
\usepackage{mathrsfs}

\usepackage[affil-it]{authblk} 
\usepackage{etoolbox}
\usepackage{lmodern}

\DeclareMathOperator\erf{erf}

\usepackage{sectsty}
\sectionfont{\fontsize{10}{15}\selectfont}

\title{\Large{Formation of Rocky Super-Earths From A Narrow Ring of Planetesimals}}
\author{\normalsize{Konstantin Batygin$^1$ \& Alessandro Morbidelli$^2$}}
\affil{\normalsize{$^1$Division of Geological and Planetary Sciences California Institute of Technology, Pasadena, CA 91125, USA | \texttt{kbatygin@gps.caltech.edu} \\ $^2$Laboratoire Lagrange, Universit\'e Cote d'Azur, CNRS, Observatoire de la Cote d'Azur, Nice, France | \texttt{alessandro.morbidelli@oca.eu}}}

\usepackage{amsmath}
\usepackage{amssymb}
\usepackage{empheq}
\usepackage{mathrsfs}
\usepackage{xcolor}

\date{}				
\usepackage{amssymb}

\begin{document}



\maketitle

\textbf{The formation of super-Earths, the most abundant planets in the Galaxy, remains elusive. These planets have masses that typically exceed that of the Earth by a factor of a few; appear to be predominantly rocky, although often surrounded by H/He atmospheres; and frequently occur in multiples. Moreover, planets that encircle the same star tend to have similar masses and radii, whereas those belonging to different systems exhibit remarkable overall diversity. Here, we advance a theoretical picture for rocky planet formation that satisfies the aforementioned constraints: building upon recent work — which demonstrates that planetesimals can form rapidly at discrete locations in the disk — we propose that super-Earths originate inside rings of silicate-rich planetesimals at approximately $\sim$1AU. Within the context of this picture, we show that planets grow primarily through pairwise collisions among rocky planetesimals, until they achieve terminal masses that are regulated by isolation and orbital migration. We quantify our model with numerical simulations and demonstrate that our synthetic planetary systems bear a close resemblance to compact, multi-resonant progenitors of the observed population of short-period extrasolar planets. Our results thus indicate that the absence of short-period super-Earths within the solar system can simply be attributed to the comparatively low mass of the primordial planetesimal ring within the protosolar nebula.}


\paragraph{Introduction.} It has long been known that the genesis of planets begins through the coalescence of solids within protoplanetary nebulae, and models of planet formation have traditionally assumed that dust within circumstellar disks is smoothly distributed. Despite being common, this simplifying assumption may be unfounded. Several lines of evidence have recently been marshaled in support of the notion that rather than arising from a smooth gradient of solids, planetesimal formation unfolds in a small number of discrete rings \cite{2014ApJ...780...53C,2016A&A...594A.105D,2022NatAs...6...72M,2022MNRAS.510.5486C,2021Sci...371..365L, 2021NatAs...6..357I}. In this vein, the work reported in ref. \cite{2022NatAs...6...72M} proposes that protoplanetary nebulae generally originate as decretion disks that spread radially from tenths of an AU, facilitating the condensation of outward-diffusing silicate vapor into rocky dust grains at the disk’s primordial silicate sublimation-line. Importantly, this process naturally leads to the formation of rocky planetesimals at a stellocentric distance comparable to the Earth’s orbital radius (as well as the generation of more distant icy bodies close to Jupiter’s present-day orbit) through gravito-hydrodynamic instabilities \cite{2005ApJ...620..459Y,2007ApJ...662..627J}. Such a model further yields a self-consistent explanation for the isotopic dichotomy of carbonaceous and non-carbonaceous iron meteorites, as well as the physical origins of the solar system’s broader architecture \cite{2022NatAs...6...72M,2021NatAs...6..357I}.

\begin{figure}[t!]
\centering
\includegraphics[width=\columnwidth]{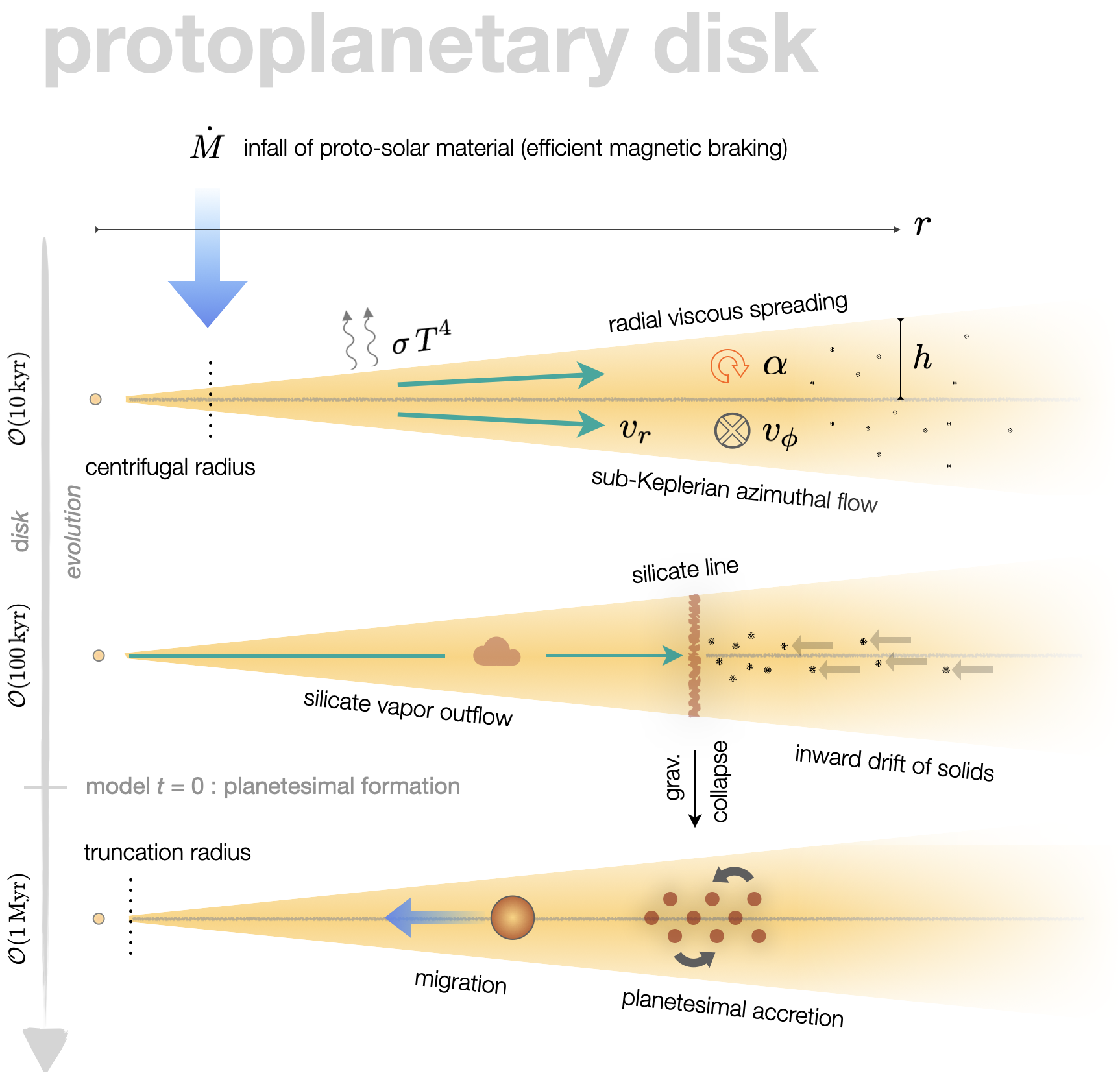}
\caption{\small{Schematic diagram of the planet formation scenario considered in this work. Strong magnetic braking during the infall phase implies that the protoplanetary nebula originates as a decretion disk that viscously spreads outward from a few tenths of an AU, carrying minute dust grains to large stellocentric distances (top panel). Beyond the condensation front, grains grow and begin to drift inwards. Accordingly, a balance between the radial outflow and sub-Keplerian azimuthal rotation of the gas leads to the accumulation of rocky grains at the silicate sublimation line, at an orbital distance of $r\sim1\,$AU (middle panel). The envisioned picture requires the satisfaction of three criteria \cite{2022NatAs...6...72M}: a small (sub-AU) centrifugal radius for the infalling material, phase-transitions of silicate species that facilitate significant changes in characteristic particle radii across one or more sublimation front(s), and a sufficiently quiescent disk for radial particle pile-up and vertical settling to occur. While the cumulative mass of the accrued dust ring is regulated by both metallicity as well as the vigor of disk turbulence, given nominal parameters, the silicate annulus can readily reach a mass on the order of tens of Earth masses. Gravito-hydrodynamic instabilities facilitate the conversion of $\sim1\,$mm dust into $\sim100\,$km planetesimals, which subsequently merge to generate multi-$M_{\oplus}$ objects. As planets deplete their local supply of solids and grow massive enough to experience substantial disk-driven migration, they exit the planetesimal feeding zone and undergo orbital decay (bottom panel), which terminates when they reach the inner edge of the protoplanetary disk. Importantly, when planets exit the planetesimal ring, their accretion stalls. Thus, the terminal mass of super-Earth type planets is approximately set by the balance between accretion and migration timescales, yielding a natural propensity towards intra-system uniformity.}}
\label{F:F1}
\end{figure} 

Within the framework of the aforementioned disk model, the mass budget of silicate material that forms at the rock-line is distinctively variable (Extended Data Figure). That is, depending on the specific combination of disk viscosity and metallicity, the cumulative mass of rocky planetesimals entrained within the $r\sim1\,$AU silicate annulus can readily reach tens of Earth masses (although we note that it can also be null if the threshold for planetesimal formation through gravitational collapse is not met). Moreover, numerical modeling indicates that planetesimal formation is expected to occur over a relatively short temporal burst, such that dust is incorporated into planetary building blocks over a timescale of $\sim10^5$ years. 

Adopting the ringed planetesimal formation paradigm as a platform, a key goal of our work is to consider the possibility that a typical system of extrasolar super-Earths originates within such a radially confined annulus of rocky material. As we describe below, the process of planetary conglomeration within a narrow ring of silicate-rich planetesimals naturally yields a characteristic multi-$M_{\oplus}$ mass scale of the resulting planets, and the simultaneous operation of accretion and orbital migration regulates the emergence of uniformity among the growing planetary embryos (Fig. \ref{F:F1}).

\paragraph{Results.} The starting point of our calculation corresponds to the epoch of large-scale planetesimal formation within a protoplanetary disk. For definitiveness, here we adopt disk conditions derived from the simulations reported in ref. \cite{2022NatAs...6...72M}, although we note that for the purposes of our calculations, any ringed planetesimal formation scenario is likely to lead to similar results. Our fiducial disk model is initialized with a gas surface density of $\Sigma_0=2500\,$g/cm$^2$ at $1\,$AU, a corresponding peak dust surface density of $\Sigma_{\bullet} = 500\,$g/cm$^2$, and a dust grain radius of $s_{\bullet}=1\,$mm, consistent with fragmentation-limited growth \cite{2010A&A...513A..56G}. Owing primarily to viscous energy dissipation, the disk maintains an appreciable aspect ratio of $h/r\sim0.05$ throughout the planetesimal formation epoch. While the gas surface density is taken to dissipate exponentially with a time-constant of $\tau_{\rm{disk}}=1.5\,$Myr, the dust surface density decays much more rapidly, owing to the fact that pebbles get incorporated into a $M_{\rm{ring}}\sim20\,M_{\oplus}$ planetesimal swarm over a $\sim10^5$ year timescale. The specific functional parameterizations of these quantities are delineated in the Methods section.

\begin{figure*}[t!]
\centering
\includegraphics[width=\textwidth]{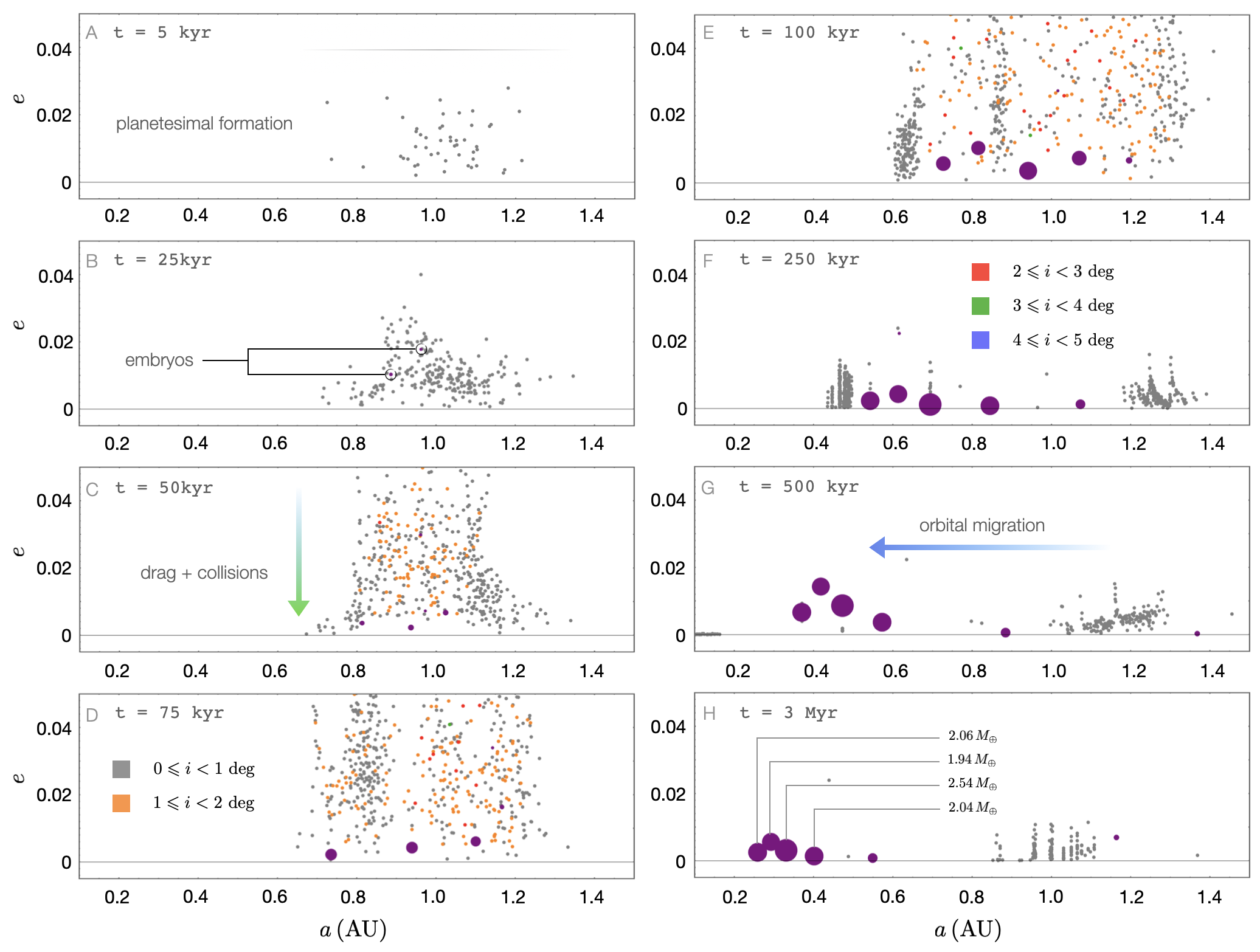}
\caption{\small{The formation sequence of a mass-uniform exoplanetary system. Over the course of the first $100{,}000$ years (panels A-E), $\mathcal{D}=100\,$km super-planetesimals (gray, orange, red, green, and blue points, labeled according to their inclinations) and lunar-mass planetary embryos (purple circles) -- comprising $M_{\rm{ring}}\approx20\,M_{\oplus}$ in total -- are gradually introduced into the simulation domain. These objects originate with eccentricities and inclinations of $\langle e \rangle \sim \langle i \rangle \sim 0.01$, across a radial range spanned by the horizontal line shown in panel A. Growth of planetary embryos is driven primarily by accretion of planetesimals, with aerodynamic drag and collisional damping facilitating enhanced gravitational focusing (panels C-E). Injection of new material into the system terminates at the $t=10^{5}$ year mark (panel E), and over the course of the following few hundred thousand years, multi-Earth-mass planets emerge, with the conglomeration process largely completed within the first $0.5\,$Myr (panel F, G). Over the course of the remaining lifetime of the disk, the formed planets migrate inwards, locking into a mass-uniform multi-resonant chain (panels H). Recent work \cite{2017MNRAS.470.1750I,2022arXiv220300801G} has shown that tightly packed multi-resonant planetary configurations serve as ideal initial conditions for reproducing both the period ratio distribution of observed extrasolar planets as well as their inferred degree of mass-uniformity.}}
\label{F:F2}
\end{figure*} 

As clouds of dust within the $r\sim1\,$AU silicate ring consolidate into planetary building blocks by means of gravitational collapse, their continued growth can proceed through two distinct channels: pairwise mergers among planetesimals and pebble accretion. The efficiency of planetesimal accretion is controlled by the extent to which gravitational focusing can increase the collisional cross-section of protoplanetary embryos. Pebble accretion, on the other hand, depends critically on whether the capture of dust proceeds in the 2D or 3D regimes — a determination that is sensitive to the characteristic size of dust particles. Generically speaking, the process of collisional fragmentation inhibits the growth of silicate grains beyond the millimeter-scale within protoplanetary disks, ensuring that even in relatively quiescent nebulae, turbulent stirring can maintain the dust sub-disk’s aspect ratio at an inflated level \cite{BM22}. Correspondingly, pebble accretion proceeds in the comparatively inefficient 3D regime, contributing very little to the planetary conglomeration process during the planetesimal formation epoch. We further find that leftover dust that is not incorporated into planetary building blocks through gravito-hydrodynamic instabilities, rapidly flows away from the planetesimal ring as the nebula matures into an accretion disk, and our estimates (see Methods section 6) indicate that any auxiliary exterior flux of pebbles plays a negligible role in driving the formation of rocky super-Earths (we confirm these analytic expectations with numerical simulations below).

\subparagraph{Analytical Estimates.} In contrast with the relative inefficiency of pebble accretion in the inner regions of a protoplanetary disk, the efficacy of planetesimal accretion within a narrow annulus of rocky planetesimals is strongly enhanced. The reasons for this are two-fold: first, by concentrating tens of Earth masses of solids into a radially confined ring of planetesimals, the rate of collisions among the constituent bodies is strongly amplified. Second, the combined action of aerodynamic drag and inelastic collisions among planetesimals constitutes a fast-acting damping mechanism for the planetesimal velocity dispersion, magnifying the effect of gravitational focusing. In this regime, the associated mass-accretion rate of a planetary embryo can be deduced from a $n-\sigma-v$ relation, and the result is well-known \cite{1993ARA&A..31..129L,2000Icar..143...15K}: $\dot{M}\sim\Sigma_{\rm{pl}}\,\pi\,R^2\,\Omega\,(1+\Theta)$, where $\Sigma_{\rm{pl}}\sim\Sigma_{\bullet}$ is the planetesimal surface density, $\Omega$ is the orbital frequency, and $\Theta=(v_{\rm{esc}}/\langle v \rangle_{\rm{pl}})^2$ is the Safronov number (i.e., the ratio of the square of the escape velocity to the square of the planetesimal velocity dispersion \cite{1969edo..book.....S}). Moreover, under the simplifying assumption of strong and time-invariant gravitational focusing, it is straightforward to show that a crude estimate for the timescale for an Earth-mass body to emerge within the ring of rocky planetesimals is given by $\mathcal{T}_{\oplus}\sim\bar{\rho}\,R_{\oplus}/(\Sigma_{\rm{pl}}\,\Omega\,\Theta)$ \cite{2020ApJ...894..143B}, where $\bar{\rho}\sim3\,$g/cc is the embryo's density. If we adopt the fiducial parameters of our model and assume that the escape velocity of the planetary embryo exceeds the planetesimal velocity dispersion by a factor of a few (corresponding to $\Theta\sim10$), $\mathcal{T}_{\oplus}$ can be as short as $\sim10^5\,$years.

No matter the dominant growth mode, planetary accretion cannot proceed without bounds. For the problem at hand, two distinct processes constitute natural termination mechanisms for planetary conglomeration, the first being isolation. Isolation occurs due to the depletion of planetesimals from the local feeding zone of the embryo. The expression for the isolation scale is easily obtained by equating the cumulative mass of planetesimals within the feeding zone (approximately two Hill radii within a heavily dissipated disk) and the planetary mass itself, to yield $M\sim8\,\pi^{3/2}\,\Sigma_{\rm{pl}}^{3/2}\,r^3/\sqrt{3\,M_{\star}}$. Given our fiducial parameters, the isolation mass within the planetesimal ring evaluates to $M\sim3\,M_{\oplus}$.

A second growth-limiting process that ensues in our model is gas-driven orbital migration. As a growing planet becomes massive enough to raise a substantial wake within the gaseous nebula, the gravitational back-reaction of the wake upon the planet drives an exchange of energy and angular momentum between the planet and the disk, which in turn expels the planet from the planetesimal ring altogether. Thus, within the framework of our theoretical picture, an approximate equivalence between the mass doubling timescale $\mathcal{T}_{\rm{mass}}\sim3\,M^{1/3}\,\bar{\rho}^{2/3}/ (\Sigma_{\rm{pl}} \, \Omega \, \Theta)$ and the migration timescale \cite{2002ApJ...565.1257T} $\mathcal{T}_{\rm{mig}}\sim(4/\Omega)(M_\star/M)(M_\star/\Sigma_0\,r^2)(h/r)^2$ yields an estimate for the mass of planets that are expected to emerge from the rocky annulus of planetesimals. Auspiciously, for the aforementioned nominal parameters, the planetary mass scale that comes out from this relation also evaluates to $M\sim3\,M_{\oplus}$.

\begin{figure}[t!]
\centering
\includegraphics[width=\columnwidth]{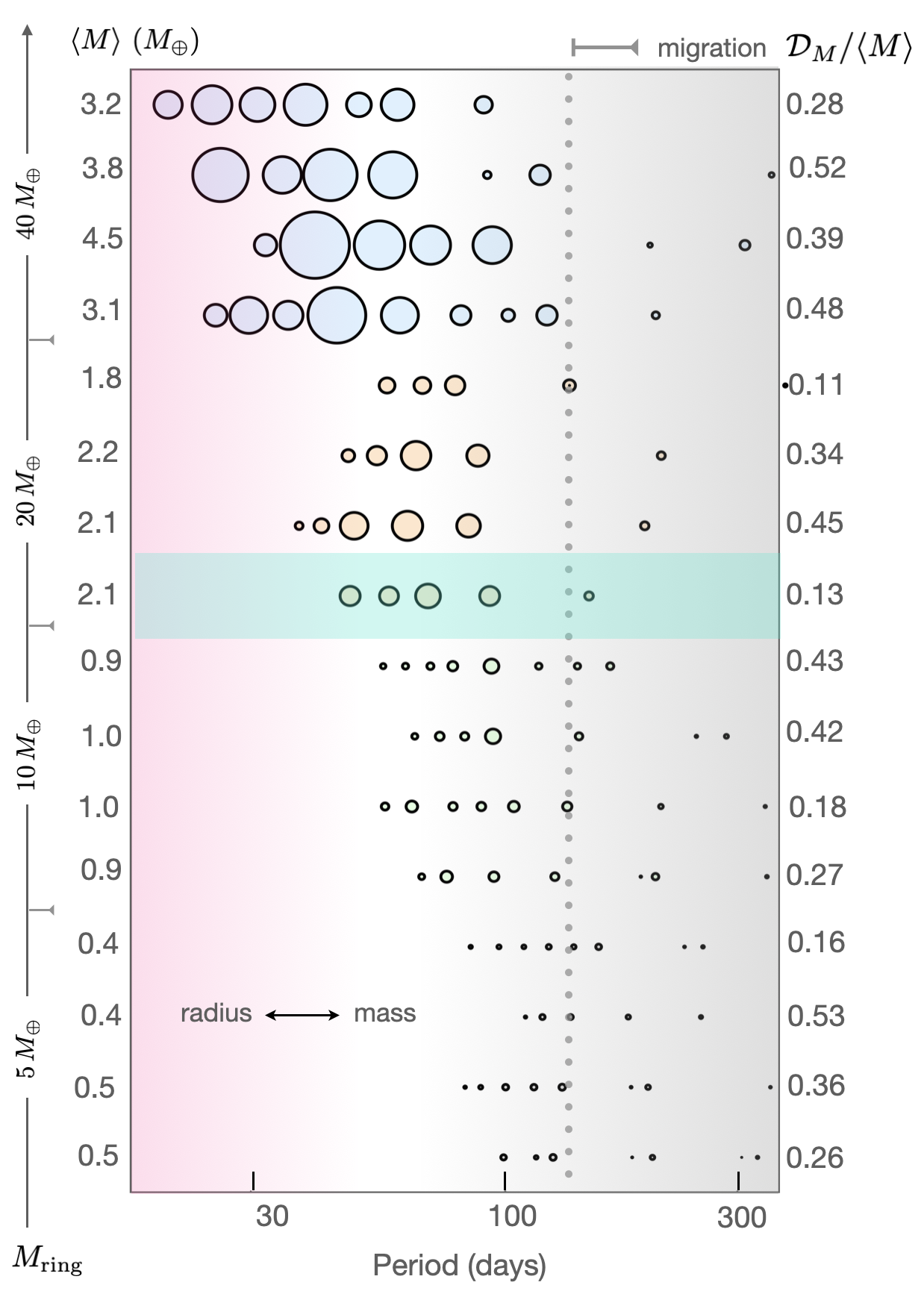}
\caption{\small{Architectures of exoplanetary systems at time of disk-dispersal, generated within the framework of our model. The formation and evolution of the system highlighted with a light-blue-green rectangle is depicted in Figure 2. As the mass of the planetesimal ring is increased from $M_{\rm{ring}}=5\,M_{\oplus}$ to $40\,M_{\oplus}$, both the number of embryos that achieve the planetary mass-scale, as well as their average mass itself increase. Within the numerical model, radial migration is taken to smoothly terminate at $r=0.5\,$AU across a characteristic length-scale of $\pm0.1\,$AU (see Methods section 5). This stellocentric distance marked by a dotted line on the Figure, and also represents the boundary between short-period progenitors to the observed population of Super-Earths and low-mass objects that remain stranded close to their formation site. By and large, these ($r<0.5\,$AU) synthetic planetary systems adhere to a pattern of intra-system mass uniformity with the normalized mass dispersion, $\mathcal{D}_{M}/\langle M\rangle$, that systematically reaches values comparable to, or smaller than, the observed value of $(\mathcal{D}_{M}/\langle M\rangle)_{\rm{data}}\approx0.48.$ Cumulatively, these results explain how planetary systems can emerge with a broad diversity of masses while retaining an unexpectedly high degree of self-uniformity.}}
\label{F:F3}
\end{figure} 

Because the isolation and accretion-migration terminal mass scales are similar, the process of planetary conglomeration is unlikely to depend sensitively on the detailed character of type-I torques, and the envisioned fiducial picture is expected to hold even in a scenario where inward migration is initially suppressed or even directed outward. And while these mass-limiting mechanisms operate simultaneously, it is nevertheless important to note that they scale differently with the planetesimal surface density. Crucially, this scaling is super-linear ($M\propto\Sigma_{\rm{pl}}^{3/2}$) and sub-linear ($M\propto\Sigma_{\rm{pl}}^{3/4}$) for isolation and migration-regulated growth respectively, meaning that migration is expected to act as the primary accretion-quenching mechanism for massive systems (that yield $M\gtrsim3\,M_{\oplus}$ planets), while isolation regulates the formation of planets in lower-mass rings.

Beyond yielding a mass-scale that is broadly consistent with the observational sample of extrasolar super-Earths \cite{a1,a2}, the scenario described above entails the emergence of a pattern of intra-system uniformity among the forming bodies \cite{2018AJ....155...48W,2017ApJ...849L..33M,2020MNRAS.493.5520A}. That is, while parameters such as the surface density, disk aspect ratio, etc., may differ from system to system, the masses of planets that form within a given disk are  likely to be similar, since they  are largely determined by the isolation and accretion-migration relations. We further remark that this paradigm is not specific to circumstellar nebulae: the standard model for the formation of giant planet satellites \cite{2002AJ....124.3404C,2020ApJ...894..143B} follows an analogous narrative, suggesting that architectural similarities between extrasolar multi-planet systems and the Galilean moons are not coincidental.

\subparagraph{Numerical Simulations.} The analytical estimates quoted above provide a useful reference point for the planetary growth sequence that is expected to ensue within a ringed disk. Nevertheless, the chaotic intricacy of planet formation cannot be captured with such considerations alone. Accordingly, we have simulated the formation of Super-Earths within an annulus of rocky planetesimals employing a full-fledged numerical model. Our simulations build upon an N-body framework and augment the self-consistent treatment of gravitational dynamics of planetesimals and planetary embryos with the effects of aerodynamic drag exerted on planetesimals by the gas, collisional damping, disk-driven (type-I) migration, as well as pebble accretion, in a parameterized manner. The details of our implementation are provided in the Methods section.

In our numerical experiments, the ring of rocky planetesimals is modeled as a population of $N_{\rm{pl}}=1000$ super-particles that form a Gaussian distribution of solid material centered on $1\,$AU, with an initial radial spread of $\Delta_{r}=0.1\,$AU. Along with $N_{\rm{emb}}=10$ lunar-mass planetary embryos, the planetesimals were introduced into the simulation gradually over a course of $10^5$ years, while the ambient surface density of solid dust is reduced in concert. Our numerical experiments followed the conventional ``big-small" categorization of bodies, wherein planetesimals could accrete onto the embryos but not onto each other. We further note that although the total number of particles in our simulations is limited by computational cost, we have also found that increasing the number of planetary seeds beyond ten does not yield materially different results, and can even diminish the realism of our simulations by over-exaggerating the effects of dynamical heating (see the Methods section for a discussion). Cumulatively, we carried out our calculations over a timespan of $3\,$Myr, in agreement with typical lifetimes of protoplanetary disks.

All in all, our numerical experiments follow a similar narrative to that outlined by the analytic estimates, and a typical formation sequence observed in our fiducial simulation suite is depicted in Fig. \ref{F:F2}. As dust is converted into planetary building blocks over the course of the first one hundred thousand years, planetary embryos experience rapid growth. Simultaneously, the initially narrow ring of solid material spreads radially due to gravitational stirring within the system. As such, a small number of separated massive objects emerge even before the epoch of large-scale planetesimal formation is complete. Over the following $\sim10^5$ years, a chaotic phase of impacts ensues, generating super-Earth-class objects. As collisional accretion wanes and the embryos dynamically isolate themselves within the planetesimal swarm, the planets enter a prolonged period of inward migration. In due course, the planetary orbits lock into a multi-resonant chain and stabilize near the disk's assumed inner edge at radii spanning tenths of an AU, where they eventually become observable by photometric surveys.

\paragraph{Discussion.} The degree of intra-system mass uniformity within our modelized planetary systems is keenly reminiscent of that observed in the data. As a specific example, the final results of the simulation shown in Fig. \ref{F:F2} are characterized by a normalized mass dispersion of $\mathcal{D}_M/\langle M \rangle \approx 0.13$. We have also carried out a variant of the same numerical experiment where the phase of long-range inward migration is delayed until after the principal phase of planetary growth is complete and obtained comparable average mass of $\langle M\rangle\approx 2\,M_{\oplus}$ and a normalized dispersion of $\mathcal{D}_M/\langle M \rangle \approx 0.42$. While these numbers are low, they not anomalous: a normalized mass dispersion of $\mathcal{D}_M/\langle M\rangle \lesssim0.5$ is an expected outcome of our proposed formation scenario. To this end, we have run a series of twelve numerical experiments akin to that depicted in Fig. \ref{F:F2}, each time randomizing the initial conditions and varying the mass of the planetesimal ring from $5$ to $40\,M_{\oplus}$. 

The census of the generated systems is depicted in Fig. \ref{F:F3}. Overall, these numerical experiments confirm that in $M_{\rm{ring}}\geqslant20\,M_{\oplus}$ systems, the typical planetary mass scales as the $\sim$3/4th power of the initial planetesimal surface density, as is expected from the accretion-migration timescale balance. Conversely, for planets generated within $M_{\rm{ring}}\leqslant20\,M_{\oplus}$ rings, this dependence is slightly super-linear ($\sim8/7$ index), signaling the increasingly important role of isolation for lower-mass systems. Regardless of the relevant scaling, the generated sample of synthetic systems conforms to a clear pattern of mass (and presumably radius) homogeneity. In agreement with recent observational determinations \cite{2022AJ....164...72M}, however, our model also predicts that this ``peas-in-a-pod" pattern is limited to short-period orbits and does not extend beyond $r\sim0.5\,$AU. Conversely, larger stellocentric distances are primarily occupied by stranded low-mass planetary objects. Finally, beyond reproducing the rocky composition of the planets themselves, the orbital architectures of our synthetic planetary systems are markedly resonant, and require post-nebular dynamical instabilities to generate the observed period-ratio distribution of the Galactic Planetary Census \cite{2017MNRAS.470.1750I,2022arXiv220300801G}.


We have not simulated the onset of such post-nebular instabilities here because the terminal point of our model corresponds to the epoch of nebular dissipation. Nevertheless, a broad array of dynamical pathways through which short-period resonant chains of planets can become unstable — including mass-loss through photo-evaporation \cite{2020ApJ...893...43M} as well as interactions with a fading quadrupolar moment of the host star \cite{2018AJ....155..167S} — is well-documented in the literature, and large-scale operation of such instabilities has already been shown to be likely \cite{2015ApJ...807...44P}. It is further worth noting that destabilization of resonant chains of planets yields a period ratio distribution that is indistinguishable from the data \cite{2017MNRAS.470.1750I} and recent work has demonstrated that the degradation in the planetary mass-homogeneity is generally very mild during a dynamical instability, such that systems that experience transient epochs of scattering remain fully consistent with the “peas-in-a-pod” pattern of uniformity observed in the population of extrasolar super-Earths \cite{2022arXiv220300801G}.

We conclude this work by remarking upon the connection between our exoplanet formation model and the formation of the Earth itself \cite{2021NatAs...6..357I}, as the two are indeed related within the context of our picture. Despite the innate complexity inherent to planetary accretion, the terminal outcome of our envisioned scenario is determined chiefly by the mass of the planetesimal ring, which in turn depends on a variety of the disk's physical properties, with turbulent viscosity playing a central role \cite{2022NatAs...6...72M}. This dependence is driven by the fact that the viscosity (and overall metallicity) controls the cumulative mass entrained within the population of rocky planetesimals (see Extended Data Figure). To this end, we note that beyond the masses of the planets themselves, their terminal orbital architecture is also sensitive to this parameter, since planets that do not accrete rapidly enough do not experience long-range inward migration, and remain close to their original formation site. Thus, within the framework of our model, one of the key reasons that the sun is encircled by the Earth — and not a group of more massive short-period planets — is simply that the protosolar nebula was sufficiently turbulent to inhibit the agglomeration of a more massive ring of rocky planetesimals at 1AU, which prevented the terrestrial planets from growing massive enough to migrate inwards before the nebular clock had run out. If the general predictions of our model endure, a unifying model for the origins of the Earth, the moons of Jupiter and Saturn, as well as extrasolar planets may finally lie within reach.

\paragraph{Data availability.} Ascii output files summarizing the time-series of our reference simulation (with an output interval of 1000 years, totaling 1010 files) are provided at https://www.konstantinbatygin.com/setimeseries.

\paragraph{Code availability.} This work utilizes the publically available \texttt{mercury6} code. The subroutine detailing user-defined forces is available on request from the corresponding author (K.B.).

\paragraph{Acknowledgments.} K.B. and A.M.  thank the referees for providing thorough and insightful reports. K. B. is grateful to Caltech, Observatoire de la C\^ote d'Azur, the David and Lucile Packard Foundation, and the National Science Foundation (grant number: AST 2109276) for their generous support. A. M. is grateful for support from the ERC advanced grant HolyEarth N. 101019380.

\paragraph{Author contributions.} K.B. and A.M. jointly conceived the project and collaborated on the interpretation of the results. K.B. carried out the N-body simulations and led the writing of the paper. A.M. ran particle-in-a-box simulations and contributed to writing of the manuscript.

\newpage

\Large{Extended Data Figure} \\

\begin{figure}[h!]
\centering
\includegraphics[width=\columnwidth]{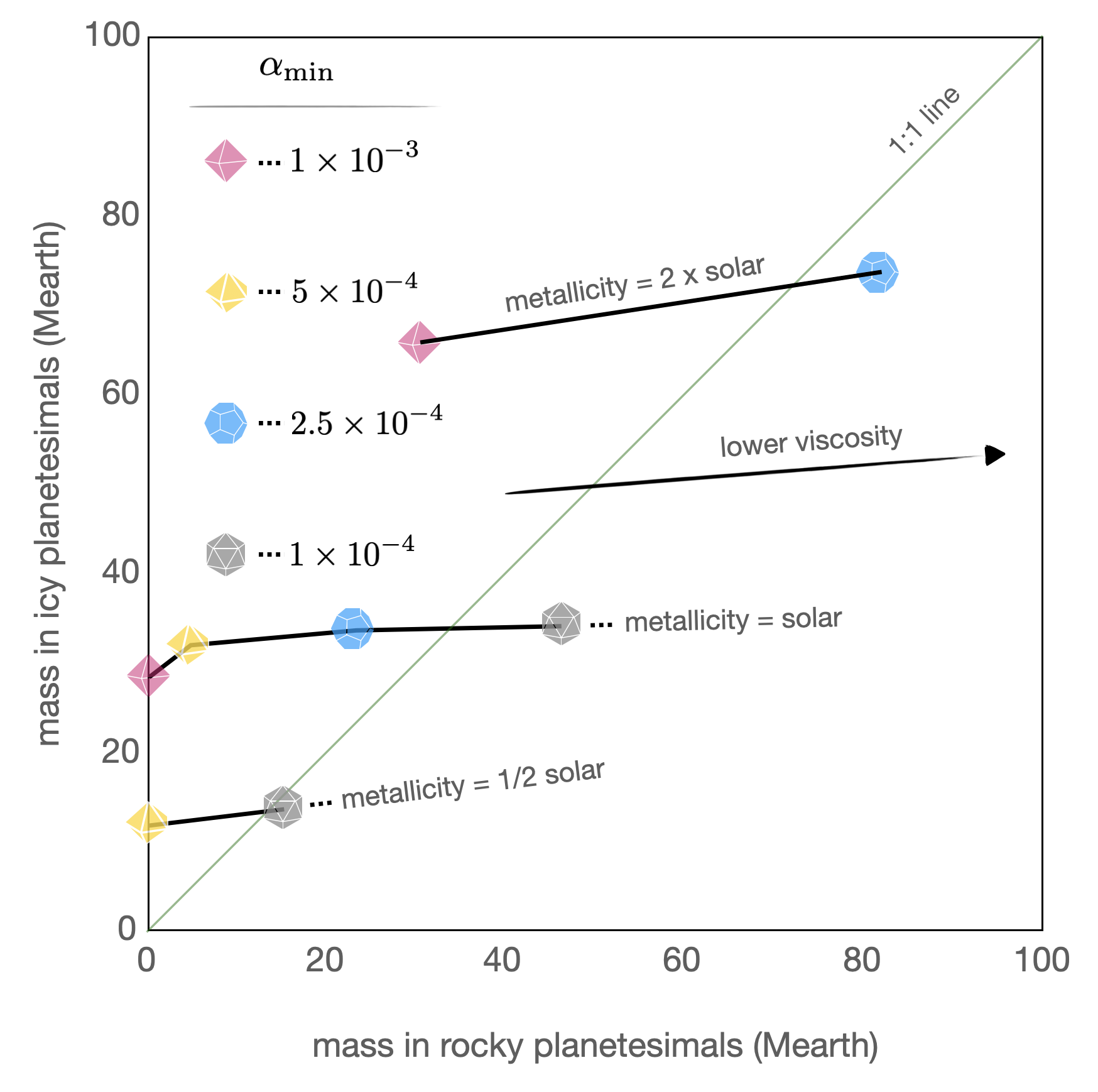}
\caption{\small{Output of the disk evolution model delineated in ref. \cite{2022NatAs...6...72M}. Owing to a hydrodynamic balance between the radial outflow and drag exerted on dust by gas, solids concentrate in the vicinity of their respective sublimation lines, forming planetesimals within discrete rocky and icy rings. Depending on the \textit{minimal} level of turbulent viscosity that the disk can attain, $\alpha_{\rm{min}}$, the mass entrained within the $r\sim1\,$AU annulus of silicate-rich material varies drastically. Importantly, for relatively quiescent nebulae, the rock ring can reach masses on the order of tens of $M_{\oplus}$, readily serving as the birthplace of super-Earths.}}
\end{figure}

\clearpage

\Large{Methods} \\

\normalsize

To quantify the concurrent processes of planetary accretion and orbital migration, we have employed the \texttt{mercury6} $N$-body simulation code \cite{1999MNRAS.304..793C}, and have augmented it to account for a series of effects that are expected to arise within protoplanetary disks. Some of these effects were modeled self-consistently, while others were implemented through \textit{ad-hoc} parameterizations for the sake of computational efficiency. Below, we describe the individual elements of our model and their physical rationale. 

\section{Protoplanetary Embryos and Super-Planetesimals: Resolution and Initial Conditions}

Following conventional practice of $N$-body calculations of planet formation, we break up our simulations into two classes of particles: fully self-gravitating planetary ``embryos" and semi-active ``planetesimals", which interact with embryos but not one-another (strictly speaking, only direct coupling between planetesimals is suppressed; indirect interactions among these particles -- that are transmitted through the reflex motion of the central body -- remain, and drive a minor but non-physical excitation of the planetesimals' velocity dispersion \cite{2020ApJ...898L..46P}). The computational cost of a given numerical experiment scales quadratically with the number of embryos ($\propto N_{\rm{emb}}^2$) and linearly with the number of planetesimals  ($\propto N_{\rm{emb}}\,N_{\rm{pl}}$). Because $N_{\rm{pl}}$ is taken to exceed $N_{\rm{emb}}$ by a large margin in typical planet formation calculations, the computational cost is primarily controlled by the product of $N_{\rm{emb}}$ and $N_{\rm{pl}}$. For this reason, we capped the planetesimal count in our simulations at $N_{\rm{pl}}=10^3$.

The initial embryo masses used in our simulations were informed by particle-in-a-box calculations of collisional growth within a $20\,M_{\oplus}$ planetesimal ring. To this end, we employed the \texttt{Boulder} code \cite{2009Icar..204..558M} to simulate the growth of protoplanetary embryos originating from $100\,$km objects, accounting for self-stirring of the planetesimal velocity dispersion, as well as collisional damping, gas drag and dynamical friction. This calculation showed the emergence of Lunar-mass planetary embryos after $\sim10{,}000$ years of evolution. 

It is important to keep in mind that even in the most numerically heavy calculations, $N_{\rm{pl}}$ is much smaller than the actual number of planetesimals that exist within protoplanetary disks. This means that each of our model planetesimals represents a ``super-particle" of mass $M_{\rm{pl}}=M_{\rm{ring}}/N_{\rm{pl}}$, and its dynamical evolution should be interpreted as a tracer of a large consortium of small bodies. Further, to simulate the inherent time-dependence of the planetesimal formation process, we injected the particles into the simulation domain at a constant rate over a span of 100,000 years, consistent with the duration of planetesimal formation epoch in the model outlined in ref. \cite{2022NatAs...6...72M}. Both the embryos and planetesimals were introduced following a Gaussian profile in the semi-major axis, centered at $\langle a\rangle=1\,$AU and a standard deviation of $\Delta_r=0.1\,$AU. Choosing a smaller value of $\Delta_r$ yielded similar results because the gravitational scattering within the planetesimal swarm facilitates a relatively rapid radial spreading of the system. The initial eccentricities and inclinations were drawn from the Rayleigh distribution with a scale parameter of $\Delta_{e}=\Delta_{i}=0.01$, while all orbital angles were uniformly randomized.

In principle, the ratio $N_{\rm{pl}}/N_{\rm{emb}}$ is an adjustable (and somewhat artibraty) parameter of the numerical model. Due to chaotic self-regulation that ensues within the planet-forming region, however, we found that the results of our calculations are only weakly dependent on the number of embryos that are injected into the annulus of rocky material during the planetesimal formation epoch. To quantify this relative insensitivity, we carried out a series of 12 simulations, where the total mass of the super-planetesimal disk was fixed at $20$ Earth masses, but the number of lunar-mass embryos was varied from $N_{\rm{emb}}=5$ to 10 to 20, running four realizations of each case. From these numerical experiments, we found that at the 300,000-year mark, the median number of proto-Super-Earths that attain a mass greater than $1M_{\oplus}$ clocked in at $3.5$, $4.5$ and $5$, for simulations with 5, 10 and 20 embryos, respectively. Furthermore, we found a broad consistency in the properties of the emergent planets with average masses of $M = 2.5 \pm 0.8\,M_{\oplus}$, $M = 2.3 \pm 1.3\,M_{\oplus}$, and $M = 2.4 \pm 1.4\,M_{\oplus}$ for the three simulation subsets.

Cumulatively, these results indicate that the dependence on the number of accreting particles saturates around  $N_{\rm{emb}}\sim10$. We did, however, find that unlike simulations with 10 embryos, those with  $N_{\rm{emb}}=20$ consistently demonstrated a pronounced (and almost certainly unphysical) difference in the velocity dispersion of the super-planetesimals and low-mass proto-planets, with the former having factor of $\sim2$ lower eccentricities on average. Consequently, we chose to adopt $N_{\rm{emb}}=10$ as a fiducial parameter in our model. With these parameters, the completion of a single run of the model on a 2.3-GHz machine required $60-80$ CPU hours.

\section{Gravitational Interactions: Integration Scheme and Time-Step}

The gravitational dynamics of our system of planetary embryos and planetesimals were solved using the hybrid Wisdom-Holman/Bulirsch-Stoer integration algorithm \cite{1991AJ....102.1528W,1992nrfa.book.....P}, implemented within the \texttt{mercury6} software package \cite{1999MNRAS.304..793C}. Because our particle swarm was initialized in the vicinity of $r\sim1\,$AU, we adopted an initial timestep of $\Delta t=10$ days. However, as disk-driven orbital evolution caused planetary orbits to decay towards their host star, we found it necessary to reduce the timestep to $\Delta t=1\,$ day at the $t=300{,}000$ year mark (and in some cases to an even lower value at later stages), to ensure that the symplectic timestep remained smaller than $\sim1/20$th of the shortest orbital period of any particle within the simulations. An adaptive time-step was used to resolve close encounters, with an inter-particle separation of $\Delta r=3\,R_{\rm{Hill}}$ marking the change-over radius for the symplectic-to-conventional integration scheme. The Bulirsch-Stoer accuracy parameter was set to $\hat{\epsilon}=10^{-8}$. Finally, the radii of embryos were computed assuming a bulk density of $3\,$g/cc, and all collisions were treated as perfect mergers. Notably, we found that in our simulations, the collision cross-section was entirely dominated by that of the embryos, meaning that our results are insensitive to the choice of the mean density of the super-planetesimals. To demonstrate this, we carried out a simulation where the physical collision radii of super-planetesimals were effectively suppressed (by choosing a corresponding density of $10^4\,$g/cc) while keeping the rest of the calculation -- including damping prescriptions -- unchanged. With such a setup, we obtained essentially identical results to our nominal simulation with planetesimal densities of $3\,$g/cc. Consequently, we adopted equal bulk densities of planetesimals and embryos, for definitiveness.

\section{Protoplanetary Nebula: Gas and Dust Profiles}

The presence of the gaseous component of the protoplanetary disk plays a key role in driving the early evolution of a forming planetary system. For definitiveness, in our work, we adopted various parameters directly derived from the ringed disk model delineated in ref. \cite{2022NatAs...6...72M} as a guide for our functional parameterizations. More specifically, we first assumed that the gas surface density followed a Mestel-like profile \cite{1963MNRAS.126..553M} that decays exponentially in time:
\begin{align}
\Sigma=\Sigma_0\,\bigg(\frac{r_0}{r}\bigg)\,\exp\big(-t/\tau_{\rm{disk}}\big),
\label{eqn:Sigma}
\end{align}
where the initial value of the surface density at $r=1\,$AU is equal to $\Sigma_0=2500\,$g/cm$^2$, and the disk decay constant is set to $\tau_{\rm{disk}}=1.5\,$Myr. Second, we adopted a constant disk aspect ratio of $h/r=0.05$ throughout the simulation. To this end, we note that while $h/r$ does in principle change in time and is in general dependent on the disk viscosity itself \cite{2015A&A...575A..28B}, this variation is not central to our numerical experiments, and is expected to only influence the results on a detailed level.

The dust component of the disk is assumed to be composed of $s_{\bullet}=1\,$mm particles, in agreement with experimentally-derived fragmentation threshold of silicate grains \cite{2010A&A...513A..56G} as well as theoretical computation of the Stokes number within the inner nebula \cite{BM22}. The dust surface density itself is envisioned to be comprised of a ``local" component -- which stems from an aerodynamically-assisted buildup of solids at the silicate sublimation front -- as well as an externally supplied flux of pebbles, $\mathcal{F}_{\rm{\bullet}}=10^{-4}\,M_{\oplus}/$yr \cite{2019A&A...627A..83L,2022arXiv220309759D}, which is facilitated by the radial drift of solids \cite{1977MNRAS.180...57W}:
\begin{align}
\Sigma_{\bullet}&=\Sigma_{\bullet\,0}\,\exp\bigg[- \bigg( \frac{r-r_0}{\Delta_r} \,\bigg)^2\, \bigg]\,\exp\bigg[- \bigg( \frac{t}{\tau_{\bullet}} \bigg)^2 \, \bigg] \nonumber \\
&+\frac{\mathcal{F}_{\rm{\bullet}}}{4\,\pi\,r\,v_{\rm{kep}}\,\eta\,\rm{St}}\,\exp\bigg[- \frac{t}{\tau_{\rm{disk}}} \bigg].
\label{eqn:Sigmadust}
\end{align}
In the above expression, $\Sigma_{\bullet\,0} = 500\,$g/cm$^2$, $\tau_{\bullet}=10^5\,$yr is the dust depletion timescale and the functional form of the leading term was chosen to adequately approximate the spatial and temporal profiles of the $r\sim1\,$AU dust ring modeled in ref. \cite{2022NatAs...6...72M}. 

Strictly speaking, $\mathcal{F}_{\bullet}$ represents the \textit{maximal} pebble flux of the nebula, and it would have been appropriate to reduce this value in accordance with sequestration of material at the rock and ice sublimation lines, to account for the finite supply of solids in the disk. As our calculations show, however, the process of pebble accretion is highly inefficient in the inner disk, so lowering the pebble flux would merely diminish an already-negligible effect. For the adopted surface density profile, the sub-Keplerian factor $\eta\approx(3/2)(h/r)^2=3.75\times10^{-3}$. Additionally, in the Epstein drag regime -- applicable for the problem at hand -- the Stokes number takes the form:
\begin{align}
\mathrm{St}=\sqrt\frac{\pi}{8}\frac{\rho_{\bullet}}{\rho}\frac{s_{\bullet}}{c_{s}}\,\Omega_{\mathrm{kep}}=\frac{\pi\,s_{\bullet}\,\rho_{\bullet}}{2\,\Sigma}.
\label{eqn:Stokes}
\end{align}

\section{Planetesimal Evolution: Aerodynamic Drag and Collisional Damping}

Interactions between planetesimals and the considerably more massive gaseous component of the disk ensue primarily through aerodynamic drag \cite{1977MNRAS.180...57W}. In the high-Reynolds number regime (appropriate for planetesimals with $\mathcal{R}\gtrsim1\,$km; \cite{1993Icar..106..264M}), the relevant drag acceleration has the form \cite{1977MNRAS.180...57W,1976PThPh..56.1756A}:
\begin{align}
\vec{a}\approx(1+\xi)\,\frac{3\,\rho}{16\,\bar{\rho}\,\mathcal{R}}\,v_{\rm{rel}}\,\vec{v}_{\rm{rel}}.
\label{eqn:drag}
\end{align}
where $\vec{v}_{\rm{rel}}\approx \vec{v}-\vec{v}_{\rm{kep}}$ refers to the relative velocity between a planetesimal and gas, and $\xi\geqslant0$ is a numerical factor that can be used to mimic the effects of non-aerodynamic damping (see below). This acceleration was implemented into our model assuming that rocky planetesimals born at $r\sim1\,$AU have diameters of $\mathcal{D}=2\,\mathcal{R}=100\,$km and bulk densities of $\bar{\rho}=3\,$g/cc.

A second important dissipative effect that affects planetesimals within the context of our calculations is collisional damping. Fundamentally, this process is driven by inelastic collisions among planetesimals, which occur with a characteristic frequency \cite{2010apf..book.....A}:
\begin{align}
\frac{1}{\tau_{\rm{coll}}}\sim \frac{3\,\Sigma_{\rm{pl}}\,\Omega}{8\,\bar{\rho}\,\mathcal{R}}.
\label{eqn:coll}
\end{align}
Although it is impossible to resolve this effect self-consistently in our super-particle calculations, we can crudely account for it by assuming that its effective functional form is similar to that of aerodynamic drag. Under this assumption, we can envision modeling the consequences of collisions (at least during the first few hundred thousand years, when the planetesimals' velocity dispersion matters most) through the enhancement factor $\xi$ introduced in equation (\ref{eqn:drag}).

To quantify the relative importance of collisional damping to aerodynamic drag -- and thereby determine the value of $\xi$ -- we begin by noting that the rate of aerodynamic damping of eccentricity and inclination is given by \cite{1976PThPh..56.1756A}:
\begin{align}
&\frac{1}{\tau_{\rm{aero}}}=-\frac{1}{e}\frac{de}{dt}=-\frac{2}{i}\frac{di}{dt}= \frac{3\,\rho\,v_{\rm{kep}}}{16\,\bar{\rho}\,\mathcal{R}}\nonumber \\
&\times\bigg(\frac{5}{8}\,e^2+\frac{1}{2}\,i^2+\eta^2\bigg)^{1/2}\sim \frac{3\,\eta\,\rho\,v_{\rm{kep}}}{16\,\bar{\rho}\,\mathcal{R}}.
\label{eqn:aerodamp}
\end{align}
The value of $\xi$ can then be approximated as:
\begin{align}
\xi=\frac{\tau_{\rm{aero}}}{\tau_{\rm{coll}}}\sim\frac{\sqrt{8\,\pi}}{\eta}\frac{\Sigma_{\rm{pl}}}{\Sigma}\frac{h}{r}\sim \sqrt{\frac{1}{2\,\pi}} \frac{1}{\eta} \frac{M_{\rm{ring}}}{\Sigma\,r_0\,\Delta_r} \frac{h}{r}.
\label{eqn:xi}
\end{align}
For nominal parameters quoted above, this factor evaluates to $\xi \sim 10$. 

While we adopt this value as an upper limit in our calculations, in order to approximately capture the dependence of the collisional damping process on the planetesimal surface density, we scale $\xi$ by the ratio $(N_{\rm{pl}}(t)/N_{\rm{pl}}^{\rm{tot}})$ in our implementation, where $N_{\rm{pl}}(t)$ is the number of super-planetesimals present in the simulation at a given timestep and $N_{\rm{pl}}^{\rm{tot}}=10^3$ is the total number of super-planetesimals injected into the simulation. Importantly, this prescription is only designed to represent the relevant physics during the first $\sim2-3\times10^5$ years of the simulation -- a period when the planetesimal population remains radially concentrated and collisional damping aids the accretion process. At later times, planetesimal dynamics play a negligible role in dictating the architecture of the emergent planetary system.

\section{Planet-Disk Interactions: Type-I Orbital Migration}
Generally speaking, the interactions between the gaseous nebula and solid material are not limited to planetesimals and aerodynamic drag. As planetary embryos accrete a sufficient amount of mass to raise significant wakes in the gaseous disk, the gravitational back-reaction of the spiral density waves upon the planet drives an exchange of energy and angular momentum \cite{1997Icar..126..261W,2018haex.bookE.139N}. In a chemically inhomogeneous disk model, both the rate and the direction of the resulting planetary migration can depend sensitively on a variety of local disk properties \cite{2010ApJ...715L..68L,2014A&A...564A.135B,2018haex.bookE.139N}, as well as the planetary mass itself. In other words, even at the qualitative level, the picture of planetary migration can be rather complex in a detailed model of the nebula.

In a simpler, power-law, locally isothermal disk model (of the type we adopt in this work) however, the sense of migration is strictly inward, and the semi-major axis, eccentricity, as well as inclination damping timescales are well defined \cite{2002ApJ...565.1257T,2004ApJ...602..388T}:
\begin{align}
&\mathcal{T}_{\rm{mig}}=\frac{\gamma}{\Omega_{\rm{kep}}}\frac{M_{\star}}{M}\frac{M_{\star}}{\Sigma\,r^2}\bigg(\frac{h}{r} \bigg)^2 \nonumber \\
&\mathcal{T}_{\rm{damp}}=\frac{\mathcal{T}_{\rm{mig}}}{2} \bigg(\frac{h}{r} \bigg)^2,
\label{eqn:migtime}
\end{align}
where the dimensionless constant $\gamma\approx4$. The effects of migration were implemented into our $N$-body scheme through auxiliary accelerations of the form \cite{2000MNRAS.315..823P}:
\begin{align}
\vec{a}=-\frac{\zeta}{\mathcal{T}_{\rm{mig}}}\,\vec{v}-\frac{2}{\mathcal{T}_{\rm{damp}}}\bigg(\frac{(\vec{v}\cdot\vec{r})\,\vec{r}}{r^2} + (\vec{v}\cdot\hat{z})\,\hat{z}\bigg),
\label{eqn:amig}
\end{align}
that were only applied to the planetary embryos.

A practically important and well-known attribute of protoplanetary disks is that they do not extend all the way down to stellar surfaces, but are instead truncated by their host stars' magnetospheres \cite{1991ApJ...370L..39K}. Due to a dramatic enhancement of the corotation torque associated with a sharp surface density gradient, the migration directions reverses at the disk's inner edge, meaning that disk cavities act as effective planet traps \cite{2006ApJ...642..478M,2018SSRv..214...38P}. The resulting stalling of inward migration at $r\sim10-20\,\mathcal{R}_{\odot}$ facilitates the formation of resonant chains that span the $a\sim0.1-0.5\,$AU range \cite{2015MNRAS.451.2589B,2017MNRAS.470.1750I,2022arXiv220300801G}.

A number of distinct approaches have been employed to model this magnetospheric cavity-driven torque reversal within the framework of $N$-body simulations, including semi-major axis re-normalization \cite{2015ApJ...810..119D,2020ApJ...894..143B}, as well as \textit{ad-hoc} (e.g., sinusoidal) modifications of the disk-driven accelerations \cite{2019A&A...627A..83L, 2021A&A...650A.152I}. While our numerical experiments are not sufficiently idealized for semi-major axis rescaling to be applicable, a drawback of the latter approach is that it requires careful tuning of the damping effects, including the introduction of non-linear eccentricity dependence in $\mathcal{T}_{\rm{damp}}$, etc., to suppress unphysical excitation of the orbits.

To avoid the usual difficulties of modeling the disk's inner edge, here we opted for a simpler approach of smoothly diminishing the migration torque interior to $r_{\rm{mig}}=0.5\,$AU such that our resonant chains stabilized with their outermost member at $a\lesssim r_{\rm{mig}}$. We implemented this by choosing the following functional form form for the multiplicative constant $\zeta$ in equation (\ref{eqn:amig}):
\begin{align}
\zeta=\frac{1}{2}\,\bigg(1+\erf \bigg[ \frac{a-r_{\rm{mig}}}{r_{\rm{mig}}/10} \bigg]\bigg).
\label{eqn:zeta}
\end{align}
In practice, we found that lowering the threshold semi-major axis to $r_{\rm{mig}}=0.25\,$AU did not alter the structure of the emergent resonant chains in any appreciable manner.

\section{Planetary Growth: Pebble Accretion}
In addition to growth facilitated by pairwise collisions between planetary embryos and planetesimals, in our simulations we also modeled direct capture of solid dust by the growing protoplanets. In general, this process -- routinely referred to as \textit{pebble accretion} \cite{2010A&A...520A..43O,2012A&A...544A..32L} -- can proceed in a number of distinct physical regimes, with their relative efficiency determined by the dust particle size, disk turbulence, planetary mass, etc. 

Recent work \cite{BM22} has argued that in typical disks, pebble accretion is expected to unfold in the comparatively inefficient 3D regime interior to the water ice sublimation line. Our assumption of a fixed $s_{\bullet}=1\,$mm particle radius yields results that are consistent with this presumption. By and large, this is due to the fact that in our model, the Stokes number (given by equation \ref{eqn:Stokes}) evaluates to very small values -- e.g., $\mathrm{St}\sim 2\times10^{-4}$ at $r\sim1\,$AU. In turn, this implies that even for relatively low values of the turbulence parameter, $\alpha$, vertical settling is prohibitively inefficient, and the dust layer's thickness remains comparable to that of the scale height of the gaseous nebula \cite{1995Icar..114..237D}:
\begin{align}
\frac{h_{\bullet}}{h}=\frac{1}{\sqrt{1+\mathrm{Sc}\,\mathrm{St}/\alpha}}\sim1.
\label{eqn:h/h}
\end{align}
We remark that the constant particle radius assumption can be lifted in favor of a more self-consistent theory for dust-gas coupling that accounts for turbulent stirring of dust grains and collisional fragmentation. Importantly, such a model also predicts that dust should be well-mixed with the gas in the inner regions of the disk, in agreement with the above estimate \cite{BM22}.

Because the dust component of the nebula never forms a thin sub-disk, the 2D regime of pebble accretion never ensues. However, for consistency with the model of ref. \cite{2022NatAs...6...72M}, where the value of the turbulent Schmidt number is taken to be on the order of $\mathrm{Sc}\sim10$, we take $h_{\bullet}/h=2/5$ in our model, such that $h_{\bullet}/r=0.02$. As we discuss below, our results are not sensitive to this choice.

The rate of 3D pebble accretion is given by \cite{2017ASSL..445..197O}:
\begin{align}
\dot{M}_{\rm{3D}}&=6\,\pi\,\mathrm{St}\,R_{\rm{Hill}}^3 \,\Omega\,\bigg( \frac{\Sigma_{\bullet}}{\sqrt{2\,\pi}\,h_{\bullet}} \bigg) \nonumber \\
&=\sqrt{2\,\pi}\,\mathrm{St}\,\Sigma_{\bullet}\,r^2\,\Omega\,\frac{M}{M_{\star}}\,\frac{r}{h_{\bullet}}.
\label{eqn:3DPA}
\end{align}
Although this growth mode was implemented in our simulations, we found pebble accretion to be largely inconsequential in our calculations. This can be understood as follows. 

As the most favorable scenario for dust capture, let us consider an isolated protoplanetary embryo of mass $m_0$ that accretes pebbles at $r=1\,$AU, where the primordial solid surface density is maximized. Substituting the functional form (\ref{eqn:Sigmadust}) for $\Sigma_{\bullet}$, it is straightforward to show that the accreted mass is bounded from above by:
\begin{align}
&\Delta{M}<\int_{0}^{\infty}\dot{M}_{\rm{3D}}\,dt = \frac{\pi}{\sqrt{2}}\,\mathrm{St}\,\Sigma_{\bullet\,0}\,r^2\,\Omega\,\tau_{\bullet}\,\frac{M_0}{M_{\star}}\frac{r}{h_{\bullet}}\nonumber \\
&+\frac{\mathcal{F}_{\bullet}\,\tau_{\rm{disk}}}{2\,\sqrt{2\,\pi}\,\eta}\frac{M_0}{M_{\star}}\,\frac{r}{h_{\bullet}}\approx 5.2 \, M_0,
\label{eqn:dMmax}
\end{align}
with the dominant contribution arising from the leading term (the externally supplied pebble flux accounts for less than a quarter of the accreted mass). 

This simple estimate indicates that even under the most optimistic conditions, pebble accretion can only boost the mass of an embryo by a factor of a few. Moreover, it is important to keep in mind that the efficiency of this process diminishes rapidly over the course of the first $\sim10^5$ years, as the majority of pebbles are converted into planetesimals (because of the smallness of this effect, we did not implement filtering of pebble flux in our model). Accordingly -- while we include the pebble accretion process in our simulations for completeness -- our numerical experiments indicate that a far more dominant role in facilitating planetary growth is played by the accretion of planetesimals and mergers among protoplanetary embryos.

\end{document}